\documentclass[12pt]{article}
\usepackage{a4}
\usepackage{amsmath}
\usepackage{latexsym}
\usepackage{amsfonts}
\begin{document}
\renewcommand{\thefootnote}{\fnsymbol{footnote}}
\thispagestyle{empty}

\noindent hep-th/   \hfill  March  2004 \\

\noindent \vskip3.3cm
\begin{center}

{\Large\bf Conformal Coupling of Higher Spin Gauge Fields to a
Scalar Field in $AdS_{4}$ and Generalized Weyl Invariance}
\bigskip\bigskip\bigskip

{\large Ruben Manvelyan\footnote{On leave from Yerevan Physics
Institute, e-mail: manvel@moon.yerphi.am} and Werner R\"uhl}
\medskip

{\small\it Department of Physics\\ Erwin Schr\"odinger Stra\ss e \\
Technical University of Kaiserslautern, Postfach 3049}\\
{\small\it 67653
Kaiserslautern, Germany}\\
\medskip
{\small\tt manvel,ruehl@physik.uni-kl.de}
\end{center}

\bigskip 
\begin{center}
{\sc Abstract}
\end{center}
The higher spin interaction currents for the conformally coupled
scalar in $AdS_{4}$ space for both regular and irregular boundary
condition corresponding to the free and interacting critical point
of the boundary $O(N)$ sigma model are constructed. The explicit
form of the linearized interaction of  the scalar and spin two and
four gauge fields in the $AdS_{D}$ space using Noether's procedure
for the corresponding spin two and four linearized gauge and
generalized Weyl transformations are obtained.

\noindent

\newpage
\renewcommand{\thefootnote}{\arabic{footnote}}\setcounter{footnote}{0}
\section{Introduction}
\quad In contrast to the case of usual $AdS_{5}/CFT_{4}$
correspondence \cite{Maldacena} where the strong coupling regime
of the boundary theory corresponds to the weak coupled
string/supergravity theory on the bulk, the $AdS_{4}/CFT_{3}$
correspondence of the critical $O(N)$ sigma model \cite{Klebanov}
operates at small t'Hooft coupling $\lambda$ and the corresponding
bulk theory is described as a theory of arbitrary even high spins.
So it is a theory of Fradkin-Vasiliev type \cite{Vasiliev}. This
case of $AdS/CFT$ correspondence is also of great interest because
here  dynamical considerations and calculations both in
$AdS_{d+1}$ and $CFT_{d}$ cases are essential on account of the
absence of supersymmetry and BPS arguments and because in this
case of correspondence perturbative expansions with small coupling
constants are mapped on each other. So the essential point of
$HS(4)$ and d=3 $O(N)$ sigma model correspondence is that both
conformal points of the boundary theory i.e. unstable free field
theory and critical interacting point, in the large $N$ limit
correspond to the same higher spin theory. Moreover as we have
learned from \cite{Klebanov} these two points are connected on the
boundary by a Legendre transformation which corresponds to the
different boundary condition (regular dimension $\beta=1$ or
shadow $\beta=2$) in the quantization of the bulk scalar field.
This quantization of the free scalar field in the $AdS$ with
different boundary conditions and corresponding multi-trace
deformation of the boundary theory were investigated and explored
in many papers, we will refer just to the articles \cite{WittKleb}
most interesting for us.

In this article we explore these two different boundary conditions
of the scalar field from the point of view of the linearized
interaction of the scalar field with the high spin fields in $d=4$
$AdS$ space. In the section 2 we extend our previous consideration
 \cite{Leonhardt} to the case of the $\beta=2$ boundary condition. We
show that if the $\beta=1$ case corresponds to the interaction of
$HS(4)$ gauge fields with the \emph{conformal } (traceless)
conserved higher spin currents constructed from scalar field, then
the $\beta=2$ case we can describe in non-contradictory fashion
with the $AdS/CFT$ correspondence  using the  non-conformal
double-traceless currents and gauge fields in Fronsdal's
formulation \cite{Fronsdal}. In the last two sections we
explicitly construct a linearized interaction \emph{Lagrangian} of
the conformal scalar field with the spin two and four gauge field
using Noether's procedure for \emph{gauge} and \emph{generalized
Weyl invariance} (some consideration of nonlinear gauge invariant
coupling of the scalar field on the level of equation of motion
one can find in \cite{ss}). We show that the interaction of the
scalar with the spin four Fronsdal gauge field can be constructed
in a non unique way due to the existence of the gauge invariant
combinations of gauge field itself (analogue of Ricci scalar for
higher spins). But this ambiguity can be fixed in unique fashion
by gauging the analogue of scale invariance in the higher spin
case. This is a symmetry with the local tensor parameter
permitting to gauge away the trace of a double traceless gauge
field and leading to the tracelessness of the corresponding spin
four current.

\section{Conformal and Fronsdal higher spin currents and $AdS_{4}/CFT_{3}$}
\quad In our previous article \cite{Leonhardt} we considered
coupling of the $HS(d+1)$ gauge field with a current constructed
from a scalar field in fixed $AdS_{d+1}$ background. Using the
ansatz including the $AdS_{d+1}$ curvature corrections we have
shown there that all ambiguity in the construction of a spin
$\ell$ \emph{traceless} current from the conformally coupled
scalar field reduces to the ambiguity of the set of leading
coefficients $A_{p}$ in the expansion of the current\footnote{For
the investigation of the conservation and tracelessness conditions
for general spin $\ell $ symmetric conformal current
$J^{(\ell)}_{\mu_{1}\mu_{2}\dots\mu_{\ell}}$ we contract it with
the $\ell$-fold tensor product of a vector $a^{\mu}$}
\begin{eqnarray}
  &&J^{(\ell)}(z;a) = \frac{1}{2}\sum^{\ell}_{p=0}A_{p}\left(a\nabla\right)^{\ell
  -p}\phi(z)\left(a\nabla\right)^{p}\phi(z) \nonumber\\
  &&+ \frac{a^{2}}{2}\sum^{\ell -1}_{p=1}B_{p}\left(a\nabla\right)^{\ell
  -p-1}\nabla_{\mu}\phi(z)\left(a\nabla\right)^{p-1}\nabla^{\mu}\phi(z)\label{ansatz}\\
  &&+\frac{a^{2}}{2L^{2}}\sum^{\ell -1}_{p=1}C_{p}\left(a\nabla\right)^{\ell
  -p-1}\phi(z)\left(a\nabla\right)^{p-1}\phi(z) + O(a^{4}) +
  O(\frac{1}{L^{4}})\;,\nonumber
\end{eqnarray}
where $A_{p}=A_{\ell-p}, B_{p}=B_{\ell-p}, C_{p}=C_{\ell-p}$ and
$A_{0}=1$. The tracelessness condition fixes relations between
$B_{p}, C_{p}$ and $A_{p}$ in the following way \cite{Leonhardt}
\begin{eqnarray}
  &&B_{p}=-\frac{p(\ell -p)}{(D+2\ell-4)}A_{p} ,\label{flatT} \\
  &&C_{p}=\frac{-1}{2(D+2\ell -4)}\left[s_{t}(p+1,\ell,D)A_{p+1}
  +s_{t}(\ell -p+1,\ell, D)A_{p-1}\right], \quad\label{L2T}\\
  &&s_{t}(p,\ell,D)=\frac{1}{4}p(p-1)D(D-2) +
  \frac{1}{3}p(p-1)(p-2)(\ell +2D-5) \;.
\end{eqnarray}
The unknown $A_{p}$ we can fix in two different ways: The first
possibility is to use the conservation condition for the current.
This leads to the recursion relation with  the same solution for
the $A_{p}$ coefficients \cite{Anselmi} as in the flat $D=d+1$
dimensional case
\begin{equation}\label{so1}
    A_{p}=(-1)^{p}\frac{\binom{\ell}{p}\binom{\ell +D-4}{p+\frac{D}{2}-2}}
    {\binom{\ell +D-4}{\frac{D}{2}-2}} \;.
\end{equation}
For the important  case $D=4$  this formula  simplifies to
\begin{equation}\label{4d}
 A_{p}=(-1)^{p}\binom{\ell}{p}^{2} \;.
\end{equation}
The result of our previous consideration \cite{Leonhardt} was the
following: the curvature corrections do not change
 flat space tracelessness  and conservation
 conditions between leading coefficients and therefore the
 solution (\ref{so1}) remains valid.
 The other way to fix this ambiguity is a boundary $CFT_{3}$ consideration.
 The result is \cite{Leonhardt}
 \begin{equation}\label{Acft}
    A_{p}=\frac{\mathcal{C^{\ell}}(-1)^{p}\binom{\ell}{p}}{2^{\ell}(\beta)_{p}
    (\beta)_{\ell-p}}\;.
\end{equation}
Here $\beta$ is the dimension of the scalar field of the boundary
$CFT_{3}$, and $\mathcal{C^{\ell}}$ is the normalization constant
of the three point function of two scalars and the spin $\ell$
current in $CFT_{3}$ (We define also the Pochhammer symbols $(z)_n
= \frac{\Gamma(z+n)}{\Gamma(z)}$.). The expression (\ref{Acft}),
for $\beta=1$, is in agreement with the previous one (\ref{4d})
obtained from $AdS_{4}$ consideration, if we will normalize in
(\ref{Acft}) $\mathcal{C^{\ell}}=2^{\ell}\ell!$. It means that the
$\beta=1$ point of boundary $CFT_{3}$ (free field conformal point
of $O(N)$ vector model) we can describe as a \emph{conformal}
$HS(4)$ model in $AdS_{4}$. In other words in this case we have to
operate in dual higher spin theory with the linearized interaction
\begin{equation}\label{cond}
S^{(\ell)conf}_{int}=\frac{1}{\ell}\int
d^{4}x\sqrt{g}h^{(\ell)\mu_{1}\dots\mu_{\ell}}J^{(\ell)}_{\mu_{1}\dots\mu_{\ell}}
,
\end{equation}
where the corresponding current is conserved and traceless
\begin{equation}
    J^{(\ell)\alpha}_{\alpha\mu_{2}\dots\mu_{\ell}}=0 \quad , \quad
    \nabla^{\mu_{1}}J^{(\ell)}_{\mu_{1}\mu_{2}\dots\mu_{\ell}}=0 .
\end{equation}

For $\beta=2$ we have to change the constraints imposed on
(\ref{ansatz}). For that we turn from conformal higher spins to
Fronsdal's \cite{Fronsdal} formulation where gauge fields and
currents are double traceless only
\begin{eqnarray}
  &&S^{(\ell)}_{int}=\frac{1}{\ell}\int
d^{4}x\sqrt{g}h^{(\ell)\mu_{1}\dots\mu_{\ell}}\Psi^{(\ell)}_{\mu_{1}\dots\mu_{\ell}} , \\
  &&
  h^{(\ell)\alpha\beta}_{\alpha\beta\mu_{5}\dots\mu_{\ell}}=0\quad,\quad
  \Psi^{(\ell)\alpha\beta}_{\alpha\beta\mu_{5}\dots\mu_{\ell}}=0 ,
  \\
  &&\delta_{0}
  h^{(\ell)}_{\mu_{1}\dots\mu_{\ell}}=\partial_{(\mu_{1}}\epsilon_{\mu_{2}\dots\mu_{\ell})},\quad
  \epsilon^{\alpha}_{\alpha\mu_{4}\dots\mu_{\ell}}=0,\quad
   \left[\nabla^{\mu_{1}}\Psi^{(\ell)}_{\mu_{1}\mu_{2}\dots\mu_{\ell}}\right]^{traceless}=0\label{cc}
\end{eqnarray}
and the conservation condition looks a little bit different from
the usual one due to the double-tracelessness of the gauge field
and current. Then we can realize the double-traceless current
$\Psi^{(\ell)}$ using two traceless (but not conserved) currents
$J^{(\ell)}$, $\Theta^{(\ell-2)}$  with the same dimension
$\ell+2\beta + O(\frac{1}{N})$ on the boundary \cite{Segal}. It
means that the expansions for these fields start from the
following series
\begin{eqnarray}
  J^{(\ell)}(z;a) &=& \frac{1}{2}\sum^{\ell}_{p=0}A^{\ell}_{p}\left(a\nabla\right)^{\ell
  -p}\phi(z)\left(a\nabla\right)^{p}\phi(z) + \dots  ,\label{al}\\
 \Theta^{(\ell-2)}(z;a)&=&
 \frac{1}{2}\sum^{\ell-1}_{p=1}B^{\ell-2}_{p}\left(a\nabla\right)^{\ell-1
  -p}\nabla_{\mu}\phi(z)\left(a\nabla\right)^{p-1}\nabla^{\mu}\phi(z) + \dots .\label{bl}
\end{eqnarray}
The Fronsdal field $\Psi^{(\ell)}$ we can present then as
\begin{eqnarray}\label{psi}
&&\Psi^{(\ell)}(z;a)=J^{(\ell)}(z;a)
+\frac{a^{2}}{2(D+2\ell-4)}\Theta^{(\ell-2)}(z;a) ,\\
&&Tr\Psi^{(\ell)}(z;a)=\Box_{a}\Psi^{(\ell)}(z;a)=\Theta^{(\ell-2)}(z;a)
\end{eqnarray}
The conservation condition (\ref{cc}) in this representation is
\begin{eqnarray}
 \nabla^{\mu}\frac{\partial}{\partial a^{\mu}} \Psi^{(\ell)}(z;a)=
 \frac{a^{2}}{2(D+2\ell-6)}Tr\nabla^{\mu}\frac{\partial}{\partial a^{\mu}}\Psi^{(\ell)}(z;a)
 \end{eqnarray}
or
\begin{equation}
\nabla^{\mu}\frac{\partial}{\partial a^{\mu}}
J^{(\ell)}(z;a)+\frac{(a\nabla)
   \Theta^{(\ell-2)}(z;a)}{(D+2\ell-4)} = \frac{a^{2}\nabla^{\mu}\frac{\partial}
   {\partial
   a^{\mu}}\Theta^{(\ell-2)}(z;a)}{(D+2\ell-6)(D+2\ell-4)} .
\end{equation}
From this we can read off a restriction on the coefficients in
(\ref{al}) and (\ref{bl})
\begin{equation}\label{main}
    p(D+2p-4)A^{\ell}_{p}+(\ell-p+1)(D+2\ell-2p-2)A^{\ell}_{p-1}+
    B^{\ell-2}_{p}+B^{\ell-2}_{p-1}=0 .
\end{equation}
For $D=4$ we get
\begin{equation}\label{4d1}
2p^{2}A^{\ell}_{p}+2(\ell-p+1)^{2}A^{\ell}_{p-1}+B^{\ell-2}_{p}+B^{\ell-2}_{p-1}=0
.
\end{equation}
Then after using (\ref{Acft}) for $\beta=2$ we obtain
\begin{equation}\label{rec}
    B^{\ell-2}_{p}+B^{\ell-2}_{p-1}=\frac{\mathcal{C^{\ell}}\ell!}{2^{\ell-1}}
    \frac{(-1)^{p}(\ell-2p+1)}{(p-1)!(p+1)!(\ell-p)!(\ell-p+2)!}\quad .
\end{equation}
The solution of this equation fulfilling the boundary conditions
\begin{equation}\label{bc}
    B^{\ell-2}_{0}=B^{\ell-2}_{\ell}=0
\end{equation}
is
\begin{equation}\label{bsolution}
    B^{\ell-2}_{p}=\frac{\mathcal{C^{\ell}}\ell!}{2^{\ell-1}}
    (-1)^{p}\sum^{p}_{k=1}\frac{(\ell-2k+1)}{(k-1)!(k+1)!
    (\ell-k)!(\ell-k+2)!} \quad .
\end{equation}
The latter sum can be proceeded using Pascal's formula for
binomials. The result is very elegant
\begin{equation}\label{bs}
    B^{\ell-2}_{p}=\frac{\mathcal{C^{\ell}}(-1)^{p}}{2^{\ell-1}(\ell+1)!}
    \binom{\ell}{p-1}\binom{\ell}{p+1} .
\end{equation}
So we show that in contrast to the $\beta=1$ case where the
interaction includes  the traceless conformal higher spin
currents, the $\beta=2$ boundary condition necessitates the
interaction with the double trace higher spin currents. The
connection between these two types of interaction can be described
adding local Weyl (in the spin two case) and generalized "Weyl"
invariants realizing the conformal coupling of the scalar with the
higher spin fields.

\section{Linearized spin two gauge and conformal scalar field interaction ($\ell=2$)}
\quad The well known action for the conformally coupled scalar
field in $D$ dimensions in external gravity  is
\begin{equation}\label{clag}
    S=\frac{1}{2}\int
    d^{D}z\sqrt{-G}\left[G^{\mu\nu}\nabla_{\mu}\phi\nabla_{\nu}\phi
    -\frac{(D-2)}{4(D-1)}R(G)\phi^{2}\right] .
\end{equation}
In this section we  restore the linearized form of this action in
fixed AdS background using a gauging procedure both for the gauge
and Weyl symmetry on the linearized level. We do this derivation
just for methodical reasons because the final nonlinear answer is
known (\ref{clag}). But we would like to extend this consideration
to the higher spin case and try to elaborate a linearized
construction which  works in the case $\ell=4$ where the final
answer is unknown.

We start from the massive free scalar action in the fixed AdS
external metric\footnote{We will use AdS conformal flat metric,
curvature and covariant derivatives commutation rules of the type
\begin{eqnarray}\label{ads}
&&ds^{2}=g_{\mu\nu}dz^{\mu}dz^{\nu}=\frac{L^{2}}{(z^{0})^{2}}\eta_{\mu\nu}dz^{\mu}dz^{\nu},\quad
\eta_{z^{0}z^{0}}=-1, \sqrt{-g}=\frac{1}{(z^{0})^{d+1}}\;,\nonumber\\
&&\left[\nabla_{\mu},\,\nabla_{\nu}\right]V^{\rho}_{\lambda} =
R^{\quad\,\,\rho}_{\mu\nu \sigma}V^{\sigma}_{\lambda}
  -R^{\quad\,\,\sigma}_{\mu\nu \lambda}V^{\rho}_{\sigma}\;,\\
 &&R^{\quad\,\,\rho}_{\mu\nu \lambda}=
-\frac{1}{(z^{0})^{2}}\left(\eta_{\mu\lambda}\delta^{\rho}_{\nu} -
\eta_{\nu\lambda}\delta^{\rho}_{\mu}\right)=-\frac{1}{L^{2}}
\left(g_{\mu\lambda}\delta^{\rho}_{\nu}
- g_{\nu\lambda}\delta^{\rho}_{\mu}\right) \;,\nonumber\\
 &&R_{\mu\nu}=-\frac{D-1}{(z^{0})^{2}}\eta_{\mu\nu}=-\frac{D-1}{L^{2}}g_{\mu\nu}\quad ,
\quad R=-\frac{D(D-1)}{L^{2}}\;.\nonumber
\end{eqnarray}}
\begin{equation}\label{lag1}
    S_{0}(\phi)=\frac{1}{2}\int
    d^{D}z\sqrt{-g}\left[\nabla_{\mu}\phi\nabla^{\mu}\phi
    +\lambda\phi^{2}\right] .
\end{equation}
For getting an interaction with linearized gravity using the
gauging procedure we have to variate $S_{0}$ with respect to
$\delta^{1}_{\varepsilon}\phi=\varepsilon^{\mu}(z)\nabla_{\mu}\phi$
\begin{eqnarray}
  \delta^{1}_{\varepsilon}S_{0} &=& \int d^{D}z\sqrt{-g}\nabla^{(\mu}\varepsilon^{\nu)}\left[\nabla_{\mu}\phi\nabla_{\nu}
  \phi
   -\frac{g_{\mu\nu}}{2} \left(\nabla_{\alpha}\phi\nabla^{\alpha}\phi +
   \lambda\phi^{2}\right)\right]
\end{eqnarray}
and solving (we assume that $\varepsilon^{\mu}$ and $h^{\mu\nu}$
have the same infinitesimal order) the equation
\begin{equation}\label{eq}
    \delta^{1}_{\epsilon}S_{0}(\phi) + \delta^{0}_{\varepsilon}S_{1}(\phi,h^{(2)})=0
    ,\quad
    \delta^{0}_{\varepsilon}h^{(2)}_{\mu\nu}=2\nabla_{(\mu}\varepsilon_{\nu)}
    ,
\end{equation}
we immediately find the following cubic interaction linear in the
gauge field
\begin{equation}\label{l1}
    S_{1}(\phi,h^{(2)})=\frac{1}{2}\int
    d^{D}z\sqrt{-g}h^{(2)\mu\nu}\left[-\nabla_{\mu}\phi\nabla_{\nu}\phi
    +\frac{g_{\mu\nu}}{2}
    \left(\nabla_{\mu}\phi\nabla^{\mu}\phi+\lambda\phi^{2}\right)\right]
    .
\end{equation}
Note that here we used many times partial integration which means
that  we admit that all fields or at least parameters of symmetry
are zero on the boundary, otherwise we would have to check all
symmetries taking into account some boundary terms and their
variations also. It is clear that for constructing the local
interaction on the bulk we can use partial integrations without
watching the boundary effects.

So we see that gauge invariance
\begin{equation}\label{gi}
\delta^{1}_{\varepsilon}\phi(z)=\varepsilon^{\mu}(z)\nabla_{\mu}\phi(z)
,\quad
\delta^{0}_{\varepsilon}h^{(2)}_{\mu\nu}(z)=2\nabla_{(\mu}\varepsilon_{\nu)}(z)
\end{equation}
in this linear approach does not fix the free parameter $\lambda$
and the corresponding spin two Noether current (energy-momentum
tensor)
\begin{equation}\label{j2}
    \Psi^{(2)}_{\mu\nu}(\phi,\lambda)=-\nabla_{\mu}\phi\nabla_{\nu}\phi
    +\frac{g_{\mu\nu}}{2}
    \left(\nabla_{\mu}\phi\nabla^{\mu}\phi+\lambda\phi^{2}\right)
\end{equation}
is conserved but not traceless. But we can fix this problem having
noted that there is one more gauge invariant combination of two
derivatives and one $h_{\mu\nu}$ field
\begin{equation}\label{r}
    r^{(2)}(h^{(2)}(z))=\nabla_{\mu}\nabla_{\nu}h^{(2)\mu\nu}-
    \nabla^{2}h^{(2)\mu}_{\mu}-\frac{D-1}{L^{2}}h^{(2)\mu}_{\mu} ,\quad
    \delta^{1}_{\varepsilon}r^{(2)}(h^{(2)})=0 .
\end{equation}
It is of course the linearized Ricci scalar-but at this moment it
is important for us that there is only one gauge invariant
combination of $h^{(2)}_{\mu\nu}(z)$ , two scalars $\phi(z)$ and
two derivatives
\begin{equation}\label{rss}
    \int d^{D}z\sqrt{g}r^{(2)}(h^{(2)})\phi^{2} ,
\end{equation}
which we can add to our linearized action with one more free
parameter. So finally we can write the most general gauge
invariant action in this approximation of the first order in the
gauge field
\begin{eqnarray}
  S^{GI}(\lambda,\xi,\phi,h^{(2)}) &=&\frac{1}{2}\int
    d^{D}z\sqrt{-g}\left[\nabla_{\mu}\phi\nabla^{\mu}\phi
    +\lambda\phi^{2}\right]  \nonumber\\
   &+&\frac{1}{2}\int
    d^{D}z\sqrt{-g}h^{(2)\mu\nu}\left[-\nabla_{\mu}\phi\nabla_{\nu}\phi
    +\frac{g_{\mu\nu}}{2}
    \left(\nabla_{\mu}\phi\nabla^{\mu}\phi+\lambda\phi^{2}\right)\right]\quad\nonumber\\
    &+& \xi\int d^{D}z\sqrt{-g}\left[\nabla_{\mu}\nabla_{\nu}h^{(2)\mu\nu}-
    \nabla^{2}h^{(2)\mu}_{\mu}-\frac{D-1}{L^{2}}h^{(2)\mu}_{\mu}\right]\phi^{2} .\label{giact}
\end{eqnarray}
Then we search for the additional local symmetry permitting to
remove the trace of the gauge field $h_{\mu\nu}$ and therefore
leading to the traceless conformal spin two current. The natural
choice here is of course  Weyl invariance and we will define local
Weyl transformation in linear approximation in the following way
\begin{equation}\label{weyl}
    \delta^{1}_{\sigma}\phi(z)=\Delta\sigma(z)\phi(z) , \quad
    \delta^{0}_{\sigma}h^{(2)}_{\mu\nu}(z)=2\sigma(z)g_{\mu\nu} ,
\end{equation}
where $\Delta$ is the conformal weight (one more additional free
parameter to fit) of the scalar field. The important point here is
that when we impose on the gauge invariant action (\ref{giact})
conformal (Weyl) invariance (\ref{weyl}) we obtain the condition
\begin{eqnarray}
 &&\frac{ \delta}{\delta\sigma(z)}S^{GI}(\lambda,\xi,\phi,h^{(2)}) =
 \left[\Delta\lambda+\frac{\lambda
   D}{2}-\frac{2\xi D(D-1)}{L^{2}}\right]\sigma\phi^{2}\\
   &&+\left[\Delta-1+\frac{D}{2}\right]\sigma\nabla_{\mu}\phi
 \nabla^{\mu}\phi+\left[2\xi(1-D)-\frac{\Delta}{2}\right]\nabla^{2}\sigma\phi^{2}=0
\end{eqnarray}
with the unique solution for  all free constants
\begin{equation}\label{const}
\Delta=1-\frac{D}{2} , \quad \xi=\frac{1}{8}\frac{D-2}{D-1},\quad
\lambda =\frac{D(D-2)}{4L^{2}} .
\end{equation}
So finally we come to the gauge and conformal invariant action
\begin{equation}\label{wiact}
S^{WI}(\phi,h_{\mu\nu})=S_{0}(\phi)+S^{\Psi^{(2)}}_{1}(\phi,h^{(2)})+S^{r^{(2)}}_{1}(\phi,h^{(2)})
\end{equation}
where
\begin{eqnarray}\label{wi}
  &&S_{0}(\phi) =\frac{1}{2}\int
    d^{D}z\sqrt{-g}\left[\nabla_{\mu}\phi\nabla^{\mu}\phi
    +\frac{D(D-2)}{4L^{2}}\phi^{2}\right] ,  \label{s0}\\
   &&S^{\Psi^{(2)}}_{1}(\phi,h^{(2)})=\frac{1}{2}\int
    d^{D}z\sqrt{-g}h^{(2)\mu\nu}\left[-\nabla_{\mu}\phi\nabla_{\nu}\phi
    +\frac{g_{\mu\nu}}{2}
    \left(\nabla_{\mu}\phi\nabla^{\mu}\phi+\frac{D(D-2)}{4L^{2}}\phi^{2}\right)\right]
    ,
    \quad\label{hcur}\quad\\
    &&S^{r^{(2)}}_{1}(\phi,h^{(2)})= \frac{1}{8}\frac{D-2}{D-1}\int d^{D}z\sqrt{-g}
    \left[\nabla_{\mu}\nabla_{\nu}h^{(2)\mu\nu}-
    \nabla^{2}h^{(2)\mu}_{\mu}-\frac{D-1}{L^{2}}h^{(2)\mu}_{\mu}\right]\phi^{2} ,\label{hr}
\end{eqnarray}
which is of course the linearized action (\ref{clag}) and can be
obtained from that after expansion near to the $AdS_{D}$
background $G_{\mu\nu}(z)=g_{\mu\nu}+h^{(2)}_{\mu\nu}(z)$ in the
first order on $h_{\mu\nu}^{(\ell)}$.
  \section{Solution for spin four}
\quad Now we start from action (\ref{wi})
 to apply Noether's method for the following
transformation of the scalar field with a traceless third rank
symmetric  tensor parameter
\begin{equation}\label{s4trans}
    \delta^{1}_{\epsilon}\phi=\epsilon^{\mu\nu\lambda}\nabla_{\mu}\nabla_{\nu}\nabla_{\lambda}\phi
    \quad , \quad
    \epsilon^{\alpha}_{\alpha\mu}=0
\end{equation}
First of all we have to calculate $\delta_{1}S_{0}$. For brevity
we introduce the notation (and in a similar way for any other
tensor)
\begin{equation}
    \tilde{\epsilon}^{\mu\nu}=\nabla_{\lambda}\epsilon^{\lambda\mu\nu}
    , \quad
    \tilde{\tilde{\epsilon}}^{\mu}=\nabla_{\nu}\nabla_{\lambda}\epsilon^{\nu\lambda\mu}
\end{equation}
Then after variation of (\ref{s0}) we obtain
\begin{eqnarray}
   &&\delta^{1}_{\epsilon}S_{0}(\phi) = \int dx^{4}\sqrt{-g}\left\{-\nabla^{(\alpha}
  \epsilon^{\mu\nu\lambda)} \nabla_{\mu}\nabla_{\alpha}\phi\nabla_{\nu}
   \nabla_{\lambda}\phi + \frac{3}{2}\tilde{\epsilon}^{\nu\lambda}\nabla_{\nu}\nabla_{\alpha}\phi
  \nabla_{\lambda}\nabla^{\alpha}\phi\right.\nonumber\\&& -\frac{1}{2}
  \tilde{\epsilon}^{\nu\lambda}\nabla^{2}\left(\nabla_{\nu}\phi\nabla_{\lambda}\phi\right)
  + \frac{1}{8L^{2}}\left[3D(D+2)-8\right]
  \tilde{\epsilon}^{\nu\lambda}\nabla_{\nu}\phi\nabla_{\lambda}\phi\\
  &&\left.-\nabla^{(\alpha}
  \tilde{\tilde{\epsilon}}^{\lambda)}\left[-\nabla_{\mu}\phi\nabla_{\nu}\phi
    +\frac{g_{\mu\nu}}{2}
    \left(\nabla_{\mu}\phi\nabla^{\mu}\phi+\frac{D(D-2)}{4L^{2}}
    \phi^{2}\right)\right]\right\}\nonumber
\end{eqnarray}
We see that we can introduce an interaction with the spin four
gauge field $h^{(4)}_{\mu\nu\alpha\beta}$ in the minimal way if we
will deform the transformation law for the spin two field. The
solution for the equation
\begin{equation}\label{eq1}
    \delta^{1}_{\epsilon}S_{0}(\phi)+\delta^{0}_{\epsilon}
    \left[S^{\Psi^{(2)}}_{1}(\phi,h^{(2)})+
    S^{\Psi^{(4)}}_{1}(\phi,h^{(4)})\right]=0
\end{equation}
is
\begin{eqnarray}
    &&S^{\Psi^{(4)}}_{1}(\phi,h^{(4)})=\frac{1}{4}\int dx^{4}\sqrt{-g}
    \left[h^{(4)\mu\nu\alpha\beta}
    \nabla_{\mu}\nabla_{\nu}\phi\nabla_{\alpha}\nabla_{\beta}\phi
    -3h^{(4)\alpha\mu\nu}_{\alpha}\nabla_{\mu}\nabla_{\beta}\phi
  \nabla_{\nu}\nabla^{\beta}\phi\right.\nonumber\\&& \quad\quad\quad\quad+\left.
  h^{(4)\alpha\mu\nu}_{\alpha}\nabla^{2}\left(\nabla_{\mu}\phi\nabla_{\nu}\phi\right)
  -\frac{3D(D+2)-8}{4L^{2}} h^{(4)\alpha\mu\nu}_{\alpha}\nabla_{\mu}\phi
  \nabla_{\nu}\phi\right] ,\label{ps4}\\
    &&\delta^{0}_{\epsilon}h^{(4)\mu\nu\alpha\beta}=4\nabla^{(\mu}\epsilon^{\nu\alpha\beta)} ,\quad
    \delta^{1}_{\epsilon}\phi=\epsilon^{\mu\nu\alpha}\nabla_{\mu}\nabla_{\nu}\nabla_{\alpha}\phi , \\
    &&\delta^{0}_{\epsilon}h^{(4)\alpha\mu\nu}_{\alpha}=2\tilde{\epsilon}^{\mu\nu}
    ,\quad
    \delta^{0}_{\epsilon}h^{(2)\mu\nu}=
    2\nabla^{(\mu}\tilde{\tilde{\epsilon}}^{\nu)} .
\end{eqnarray}
So we obtain the following gauged action with linearized
 interaction with both spin two and spin four gauge fields and
 linearized usual Weyl invariance
\begin{eqnarray}\label{gi4}
    && S^{GI}(\phi,h^{(2)},h^{(4)})=S^{WI}(\phi,h^{(2)})+S^{\Psi^{(4)}}_{1}(\phi,h^{(4)})\\
    &&\delta^{0}h^{(4)\mu\nu\lambda\alpha}=4\nabla^{(\mu}\epsilon^{\nu\lambda\alpha)},
    \quad\delta^{0}h^{(2)\mu\nu}=2\nabla^{(\mu}\varepsilon^{\nu)}
    + 2\nabla^{(\mu}\tilde{\tilde{\epsilon}}^{\nu)}+2\sigma g_{\mu\nu}\\
    &&\delta^{1}\phi=\varepsilon^{\mu}\nabla_{\mu}\phi +
    \epsilon^{\mu\nu\lambda}\nabla_{\mu}\nabla_{\nu}\nabla_{\lambda}\phi
    +(1-\frac{D}{2})\sigma\phi
\end{eqnarray}
where $S^{WI}(\phi,h^{(2)})$ can be read from
(\ref{wiact})-(\ref{hr}) and we note that on this linearized level
usual Weyl transformation does not affect the spin four part of
the action but the spin four gauge transformation deforms the
gauge transformation for spin two gauge field.

Now we turn to the construction of the conformal invariant
coupling of the scalar field  with the spin four gauge field in a
similar way as in the case of spin two. For this we note first
that here we can construct also the gauge invariant combination of
two derivatives and $h^{(4)\mu\nu\alpha\beta}$. This is the
following traceless symmetric second rank tensor
\begin{eqnarray}
    &&r^{(4)\alpha\beta}=\nabla_{\mu}\nabla_{\nu}h^{(4)\mu\nu\alpha\beta}
    -\nabla^{2}h^{(4)\mu\alpha\beta}_{\mu}-\nabla^{(\alpha}\nabla_{\nu}h^{(4)\beta)\mu\nu}_{\mu}
    -\frac{3(D+1)}{L^{2}}h^{(4)\alpha\beta\mu}_{\mu} ,\quad\label{r4}\\
    &&\delta^{1}_{\epsilon}r^{(4)\alpha\beta}=0 ,\quad\quad
    r^{(4)\alpha}_{\alpha}=0
\end{eqnarray}
This is the analogue of the Ricci scalar in the spin four case and
we can construct using this tensor \emph{two} additional gauge
invariant combinations of the same order.
\begin{equation}\label{r4z}
    S^{r^{(4)}}_{1}(\xi_{1},\xi_{2},\phi,h^{(4)})=\xi_{1}\int
    d^{D}z\sqrt{-g}r^{(4)\mu\nu}\nabla_{\mu}\phi\nabla_{\nu}\phi +
    \xi_{2}\int
    d^{D}z\sqrt{-g}\nabla_{\mu}\nabla_{\nu}r^{(4)\mu\nu}\phi^{2} .
\end{equation}
Then we can define the \emph{generalized} "Weyl" transformation
for the scalar and spin four gauge field with the second rank
symmetric traceless parameter $\chi^{\mu\nu}(z)$
\begin{eqnarray}
   \delta^{0}_{\chi}h^{(4)\mu\nu\alpha\beta}(z)=12\chi^{(\mu\nu}(z)g^{\alpha\beta)} ,
  \quad \delta^{1}_{\chi}\phi(z)=\tilde{\Delta}\chi^{\alpha\beta}(z)
  \nabla_{\alpha}\nabla_{\beta}\phi(z) ,
 \end{eqnarray}
where we introduced the "conformal" weight $\tilde{\Delta}$ for
the scalar field. Computing the following $\chi$ variations
\begin{eqnarray}
  &&\delta^{1}_{\chi}S_{0}(\phi)+\delta^{0}_{\chi}S_{1}^{\Psi^{(4)}}(\phi,h^{(4)})=
  \int \left\{(\tilde{\Delta}-1)\nabla^{(\alpha}\tilde{\chi}^{\beta)}
  \Psi^{(2)}_{\alpha\beta}(\phi,\frac{D(D-2)}{4L^{2}})\right.\nonumber\\
&&-(\tilde{\Delta}+\frac{3D}{2}+3)\chi^{\alpha\beta}\nabla_{\alpha}
\nabla_{\mu}\phi\nabla_{\beta}\nabla^{\mu}\phi
+\frac{\tilde{\Delta}+D+3}{2}\nabla^{2}\chi^{\alpha\beta}\nabla_{\alpha}\phi\nabla_{\beta}\phi\nonumber\\
&&\left.-\frac{1}{L^{2}} C(\tilde{\Delta},D)\chi^{\alpha\beta}
\nabla_{\alpha}\phi\nabla_{\beta}\phi +\frac{D(D-2)}{8L^{2}}
\tilde{\tilde{\chi}}\phi^{2}\right\}\sqrt{-g}d^{D}z
   ,\label{var1}\\
&&C(\tilde{\Delta},D)=(\tilde{\Delta}-1)(D-1)+\frac{\tilde{\Delta}}{4}D(D-2)+
(D+4)(\frac{3D(D+2)}{8}-1) ,\quad\label{c}\\
&&\delta^{0}_{\chi}S_{1}^{r^{(4)}}(\phi,h^{(4)})=\xi_{1}\int
\left[2D\nabla^{(\alpha}\tilde{\chi}^{\beta)}
\Psi^{(2)}_{\alpha\beta}(\phi,\frac{D(D-2)}{4L^{2}})-(D-2)\tilde{\tilde{\chi}}\nabla_{\alpha}
\phi\nabla^{\alpha}\phi
\right.\quad\quad\nonumber\\
&&-2(D+3)\nabla^{2}\chi^{\alpha\beta}\nabla_{\alpha}\phi\nabla_{\beta}\phi
\left.-\frac{2}{L^{2}}(D+3)(3D+4)\chi^{\alpha\beta}
\nabla_{\alpha}\phi\nabla_{\beta}\phi\right]\sqrt{-g}d^{D}z\nonumber\\
&&-\left[\xi_{1}\frac{D^{2}(D-2)}{4L^{2}}+\xi_{2}\frac{12(D+1)(D+2)}{L^{2}}\right]\int
d^{D}z\sqrt{-g}\tilde{\tilde{\chi}}\phi^{2}\nonumber\\&&-\xi_{2}4(D+1)\int
d^{D}z\sqrt{-g}\nabla^{2}\tilde{\tilde{\chi}}\phi^{2}\label{vars}
\end{eqnarray}
we see again that for obtaining a "Weyl" invariant interaction we
have to deform the gauge and usual Weyl transformation of the spin
two gauge field $h^{(2)}_{\mu\nu}$
\begin{equation}\label{def}
    \delta^{0}_{\chi}h^{(2)}_{\mu\nu}=
    2(1-\tilde{\Delta}-2D\xi_{1})\nabla^{(\mu}\tilde{\chi}^{\nu)}+2\xi_{1}\tilde{\tilde{\chi}}g_{\mu\nu}
\end{equation}
Then solving the symmetry condition
\begin{equation}\label{sc}
\delta^{1}_{\chi}S_{0}(\phi)+\delta^{0}_{\chi}\left(S_{1}^{\Psi^{(2)}}(\phi,h^{(2)})+S_{1}^{r^{(2)}}(\phi,h^{(2)})
+S_{1}^{\Psi^{(4)}}(\phi,h^{(4)})+S_{1}^{r^{(4)}}(\phi,h^{(4)})\right)=0
\end{equation}
we obtain again a unique solution for all three free parameters
\begin{eqnarray}
   && \tilde{\Delta}=-3-\frac{3}{2}D , \\
  && \xi_{1}=-\frac{1}{8}\frac{D}{D+3} , \\
  && \xi_{2}=\frac{1}{64}\frac{D(D-2)}{(D+1)(D+3)} .
\end{eqnarray}
Thus we constructed the linearized action for a scalar field
interacting with the spin two and four field in a conformally
invariant way
\begin{eqnarray}
  &&S^{WI}(\phi,h^{(2)},h^{(4)}) = S^{WI}(\phi,h^{(2)})+S_{1}^{\Psi^{(4)}}
  (\phi,h^{(4)})+ S_{1}^{r^{(4)}}(\phi,h^{(4)}) ,\\
  && \delta^{1}\phi=\varepsilon^{\mu}\nabla_{\mu}\phi +
    \epsilon^{\mu\nu\lambda}\nabla_{\mu}\nabla_{\nu}\nabla_{\lambda}\phi
    +\Delta\sigma\phi+\tilde{\Delta}\chi^{\mu\nu}\nabla_{\mu}\nabla_{\nu}\phi ,\\
  &&\delta^{0}h^{(2)\mu\nu}=2\nabla^{(\mu}\varepsilon^{\nu)}
    + 2\nabla^{(\mu}\tilde{\tilde{\epsilon}}^{\nu)}+2(1-\tilde{\Delta}-2D\xi_{1})
    \nabla^{(\mu}\tilde{\chi}^{\nu)}+2\sigma
    g_{\mu\nu}+2\xi_{1}\tilde{\tilde{\chi}}g_{\mu\nu} ,\quad\quad\\
    &&\delta^{0}h^{(4)\mu\nu\alpha\beta}=4\nabla^{(\mu}\epsilon^{\nu\lambda\alpha)}
    +12\chi^{(\mu\nu}g^{\alpha\beta)} .
\end{eqnarray}

This interaction has an additional local symmetry permitting to
gauge away the trace of spin two and four gauge fields. So we can
say that this is a linearized interaction for \emph{conformal
higher spin theory} of the type discussed in
\cite{Segal},\cite{Fradkin}. Unfortunately at the moment we can
present only the spin four case in a complete form. But the
general spin $\ell$ case in $AdS$ can be considered in a similar
but more complicated way and will be presented in  future
publications.
\subsection*{Acknowledgements}
\quad This work is supported in part by the German
Volkswagenstiftung. The work of R.~M. was supported by DFG
(Deutsche Forschungsgemeinschaft) and in part by the INTAS grant
\#03-51-6346 .

\end{document}